\documentclass[aps,prl,twocolumn,letterpaper,superscriptaddress,10pt, floatfix]{revtex4-2}

\usepackage{amssymb,amsthm,amsmath,amsfonts}
\usepackage{graphicx,ulem,enumerate,mathptmx}
\usepackage[pdftex,dvipsnames,usenames]{xcolor}
\usepackage[colorlinks=true,urlcolor=blue,citecolor=blue,linkcolor=blue]{hyperref}
\usepackage{braket}

\begin{document}

\title{Realization of high-fidelity unitary operations on up to 64 frequency bins}

\author{Syamsundar~De}
\affiliation{Integrated Quantum Optics Group, Institute for Photonic Quantum Systems (PhoQS), Paderborn University, 33098 Paderborn, Germany}
\affiliation{Advanced Technology Development Centre, IIT Kharagpur, Kharagpur 721302, India}

\author{Vahid~Ansari}
\affiliation{Integrated Quantum Optics Group, Institute for Photonic Quantum Systems (PhoQS), Paderborn University, 33098 Paderborn, Germany}
\affiliation{E. L. Ginzton Laboratory, Stanford University, 348 Via Pueblo Mall, Stanford, California 94305, USA}

\author{Jan~Sperling}
\affiliation{Theoretical Quantum Science, Institute for Photonic Quantum Systems (PhoQS), Paderborn University, 33098 Paderborn, Germany}

\author{Sonja~Barkhofen}
\affiliation{Integrated Quantum Optics Group, Institute for Photonic Quantum Systems (PhoQS), Paderborn University, 33098 Paderborn, Germany}

\author{Benjamin~Brecht}\email{benjamin.brecht@upb.de}
\affiliation{Integrated Quantum Optics Group, Institute for Photonic Quantum Systems (PhoQS), Paderborn University, 33098 Paderborn, Germany}

\author{Christine~Silberhorn}
\affiliation{Integrated Quantum Optics Group, Institute for Photonic Quantum Systems (PhoQS), Paderborn University, 33098 Paderborn, Germany}

\date{\today}

\begin{abstract}
    The ability to apply user-chosen large-scale unitary operations with high fidelity to a quantum state is key to realizing future photonic quantum technologies. 
    Here, we realize the implementation of programmable unitary operations on up to 64 frequency-bin modes. 
    To benchmark the performance of our system, we probe different quantum walk unitary operations, in particular Grover walks on four-dimensional hypercubes with similarities exceeding 95\% and quantum walks with 400 steps on circles and finite lines with similarities of 98\%. 
    Our results open a new path towards implementing high-quality unitary operations, which can form the basis for applications in complex tasks, such as Gaussian boson sampling. 
\end{abstract}

\maketitle

\paragraph{Introduction.\textemdash\hspace*{-2ex}}
    Linear optical networks play a key role for many photonic quantum technologies. 
    They underlie ongoing efforts in realizing photonic quantum simulations \cite{Walther:12}, communications \cite{kimble_quantum_2008}, and computations \cite{Knill:01}.
    One recent example for these technologies are boson sampling \cite{aaronson2011computational} and Gaussian boson sampling \cite{PhysRevLett.119.170501}, where one sends quantum states of light into the input modes of a large, random interferometer built from beam splitters and phase shifters\textemdash a linear optical network\textemdash and detects the photon statistics in the output modes.
    Using this framework, a quantum advantage has been successfully demonstrated on different experimental platforms \cite{doi:10.1126/science.abe8770, madsen_quantum_2022}.
    For moving forward, however, it is crucial that we are able to precisely control the exact unitary operation that is implemented by the network \cite{Burgwal:17}.
    One way to realized this, is to build the network from individual beam splitters and phase shifters whose properties can be individually set, and in fact it has been shown that this allows for implementing any arbitrary unitary operation \cite{PhysRevLett.73.58, Clements:16}.
    By following this path, we implicitly choose to encode the input modes of our linear optical network into spatial paths.
    On a first glance, this seems a rather reasonable choice as it naturally mimics the topology of integrated photonic (quantum) circuits, which promise scalability through integration (see, e.g., \cite{Flamini:18} and the references therein). 
    On a second glance, however, we are facing two major challenges: 
    firstly, when scaling to larger network size, we have to traverse more optical elements which leads to an exponential increase of the losses; 
    secondly, an $(N\times N)$ unitary operation requires on the order of $N^2$ beam splitters and phase shifters and fabrication tolerances will necessarily limit the achievable overall fidelity \cite{Burgwal:17}.

    Therefore, to guarantee both \textit{scalability} and \textit{high fidelity} simultaneously, we should consider alternative approaches for realizing large unitary operations. 
    This means that we should think about other, high-dimensional degrees of freedom as our fundamental encoding basis and then identify ways to interfere different basis elements. 
    In recent years, this has been studied for spatial modes \cite{Armstrong:12}, frequency bins \cite{Reimer:19, PhysRevLett.125.120503, Lu:22}, time bins \cite{PhysRevLett.125.213604, madsen_quantum_2022}, and broadband temporal modes \cite{Brecht15, PhysRevLett.121.220501}. 
    Of these, time-frequency encodings are particularly appealing because they are naturally compatible with single-mode fibers and, consequently, integration
    
    In this work, we introduce an innovative approach to realizing frequency-encoded linear optical networks that combines the temporal-mode framework and frequency-bin encoding. 
    This will allow us to implement programmable linear optical networks in one single waveguide.
    To this end, we expand our original temporal-mode framework \cite{Brecht15} to frequency bins (FBs), which represent the input modes of a linear optical network.
    The unitary operation of the linear optical network that is, the coherent mixing of different FBs, and the detection of desired output modes will be realized with a quantum pulse gate (QPG) followed by detection with low-cost silicon avalanche photo-diodes.
    the QPG is an all-optical element based on dispersion-engineered, guided-wave sum-frequency generation \cite{Brecht14}, which sets our approach apart from other approaches based on electro-optic modulation \cite{Reimer:19, PhysRevLett.125.120503, Lu:22}. 
    We will demonstrate that we can implement up to $(64 \times 64)$ unitary operations with high fidelity in a single step, and that we can configure both the size and structure of our unitary without physically changing the experiment. We emphasize that this includes full connectivity between arbitrarily chosen modes of the network.

\begin{figure}[t]
	\includegraphics[width=\columnwidth]{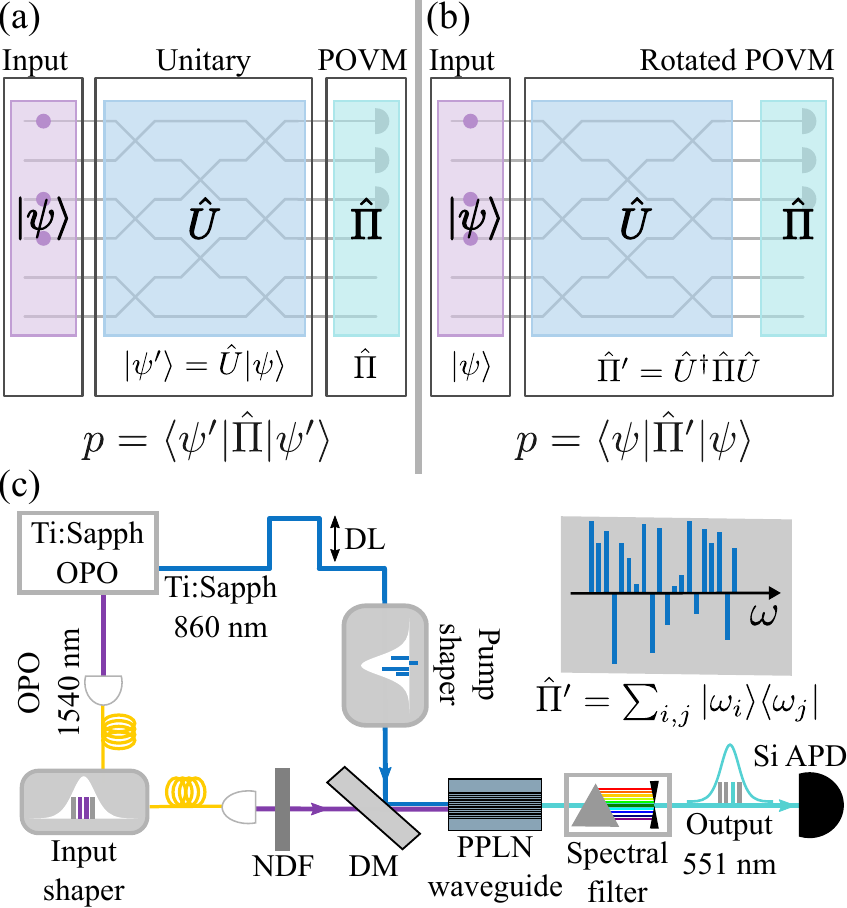}
	\caption{
	    Concept of our scheme.
	    (a) Conventional approach: a multi-mode input state $\ket{\psi}$ evolves under the action of a unitary operation $\hat U$ and is subject to a POVM $\hat\Pi$. 
	    The resulting probabilities $p$ are given by Born's rule, $p = \bra{\psi'} \hat\Pi \ket{\psi'}$. 
	    (b) Our approach: we prepare a multi-mode input state $\ket\psi$. 
        We then apply the unitary operation $\hat U$ to $\hat\Pi$ and obtain a rotated POVM $\hat\Pi'=\hat U^{\dagger} \hat\Pi \hat U$, which yields probabilities $p = \bra\psi \hat\Pi' \ket\psi = \bra{\psi'} \hat\Pi \ket{\psi'}$. 
	    (c) Schematic of our experimental setup.  
	    The inset illustrates an example of a projector onto frequency-bin superpositions.   
	    For details, see the text;
	    titanium sapphire oscillator (Ti:Sapph), optical parametric oscillator (OPO), neutral density filter (NDF), dichroic mirror (DM), delay line (DL), periodically poled lithium niobate (PPLN), silicon avalanche photo-diode (SiAPD). 
	    }\label{fig:1}
\end{figure}

\paragraph{Realizing arbitrary $(N \times N)$ unitary operations.\textemdash\hspace*{-2ex}}
    Fig. \ref{fig:1} illustrates the concept behind and implementation of our approach. 
    In a conventional linear optical system, an initial multi-mode state $\ket\psi$ is subject to a unitary evolution $\hat U$, which yields a final state $\ket{\psi'} = \hat U \ket\psi$. 
    The final state is then measured in the original mode basis, which is described by the action of a positive-operator-valued measurement (POVM) $\hat\Pi$, which acts on one or several of the modes. The resulting probabilities $p$ are characteristic for the system; see Fig. \ref{fig:1} (a). 
    In our scheme, similar to other schemes \cite{Cai:17}, the unitary operation $\hat U$ is applied to the POVM $\hat\Pi$, which yields a rotated measurement $\hat\Pi'=\hat U^\dagger \hat\Pi \hat U$. 
    This measurement projects the initial state $\ket\psi$ onto superpositions of modes and leads to the same probabilities as the conventional scheme; see Fig. \ref{fig:1} (b).

    While this technique is challenging to implement for spatial-path or time-bin encodings, projections onto arbitrary temporal-mode superposition are easily implemented with the QPG \cite{Ansari17, Ansari18prl, Ansari20}.
    The complete scheme of our setup is sketched in Fig. \ref{fig:1}(c) and more details are given in \cite{supp}. 
    A coherent light pulse, attenuated to a few-photon level, acts as the input state $\ket\psi$.
    It is carved from a broadband optical parametric oscillator (OPO) with a central wavelength of 1540~nm and a spectral intensity full-width-at-half-maximum (FWHM) of 2.8~THz with the help of a commercial wave shaper with a resolution of 10~GHz.
    We set the width of each FB to 40~GHz, with interspersed guard bands of 36~GHz.
    This binning configuration ensures a good trade-off between basis orthogonality and maximum number of accessible modes.
    The pump for the QPG is derived from an ultrafast Ti:Sapph oscillator, delivering pulses with a central wavelength of 860~nm and a spectral intensity FWHM of 2.8~THz. 
    A home-built pulse shaper with a resolution of 20~GHz enables us to control the FB structure of the pump, which is required for projecting onto desired FB-superpositions.
    A schematic representation of a programmed FB superposition is depicted in the inset of Fig. \ref{fig:1}(c) and we emphasize that different phases and amplitudes can be set for each bin, which are generally defined by the desired unitary $\hat{U}$.
    Ti:Sapph and OPO pulses are temporally overlapped via free-space delay lines and recombined on a dichroic mirror, after which they are coupled to the QPG. 
    Detecting photons at the output port of the QPG that is, at the sum-frequency (around 551~nm) with a single-photon detector implements the projection $\hat{\Pi}'$. 
    The QPG is a 35~mm-long, periodically poled lithium niobate waveguide, which was fabricated in house.
    The sum-frequency output is singled out and spectrally filtered with a bandwidth of 50~GHz to discard the phase matching sidelobes. 
    This increases the mode selectivity and hence the fidelity of the projection.
    The spectrally filtered light is then coupled to a single-mode fiber and detected with a silicon avalanche photodiode.
    From the time-integrated counts for each detection mode, we calculate the associated probabilities $\{p\}$, c.f. Fig. \ref{fig:1}(b).

\paragraph{Experimental testbed. \textemdash\hspace*{-2ex}}
    To benchmark the performance of our system we choose to implement unitary operations typically associated with quantum walk (QW) evolutions.
    QWs are well-understood and find application in the context of quantum simulations \cite{Walther:12}.
    They can also implement universal quantum computation \cite{PhysRevLett.102.180501, PhysRevA.81.042330}, if stringent conditions on scalability and fidelity can be satisfied.

    In a general discrete-time QW, a walker coherently propagates over a set of external positions $\Vec{x}$ in a \mbox{$d_x$-dimensional} space, together with $d_c$ internal coin states $c$, yielding a total dimensionality of $d=d_x\cdot d_c$.
    The joint state of the position-coin system in any given step $n$ can be expanded as $|\psi_n\rangle=\sum_{c,\Vec{x}}\psi_{n,c}(\Vec{x})|c\rangle\otimes|\Vec{x}\rangle$.
    The $n$-th step of the walk reads $|\psi_{n-1}\rangle	\mapsto  |\psi_n\rangle=\hat{U}_n|\psi_{n-1}\rangle$, where, the evolution operator decomposes as $\hat{U}_n=\hat{S}_n(\hat{C}_n\otimes\hat{I})$.
    The coin operator $\hat{C}_n$ defines the quantum coin toss and acts on the internal degree of freedom. 
    The step operator $\hat{S}_n$ shifts the walker’s position depending upon the coin state. 
    The complete evolution of a QW is given by the successive application of each step unitary operation such that
    \begin{equation}
        \ket{\psi'} = \prod_{n=1}^N \hat U_n \ket{\psi} = \hat U \ket{\psi} 
    \end{equation}
    applies, where $\ket{\psi}$ and $\ket{\psi'}$ are the initial and final states of the walker, respectively, and $\hat U$ is the $(d \times d)$ unitary describing the complete QW.

    In our experiment, FBs encode both the position and coin degrees of freedom. 
    We compute the elements of the rotated measurement basis by applying the unitary matrix $\hat U$ describing the QW to the original frequency-bin basis. 
    We then successively project onto each element of the measurement basis and collect output photons. 
    At this stage, the resulting output probabilities $\{p\}$ are still coin-resolved; 
    that is, they contain information about the external and internal state of the walker. 
    Position distributions\textemdash the common QW metric\textemdash are obtained by tracing out the internal degree of freedom, such that
    \begin{equation}
    \label{eq:rotated_measurement}
    \begin{aligned}
        P(\Vec{x})=&\sum_c \bra{\psi}\hat{\Pi}'\ket{\psi}.
    \end{aligned}
    \end{equation}

    We emphasize that we do not claim to \textit{implement} an actual QW, as we do not evolve a walker with time.
    Rather, we implement the unitary $\hat{U}$ associated with a specific QW to benchmark the performance of our approach.
    Ultimately, for quantum technologies applications that rely on the propagation of quantum light through linear optical networks and subsequent measurement, the distinction becomes irrelevant, however, as our approach yields exactly the same output statistics as a true evolution followed by a measurement in the original basis.

\begin{figure*}[t]
	\includegraphics[width=\textwidth, trim=0cm 4.5cm 0cm 0cm, clip]{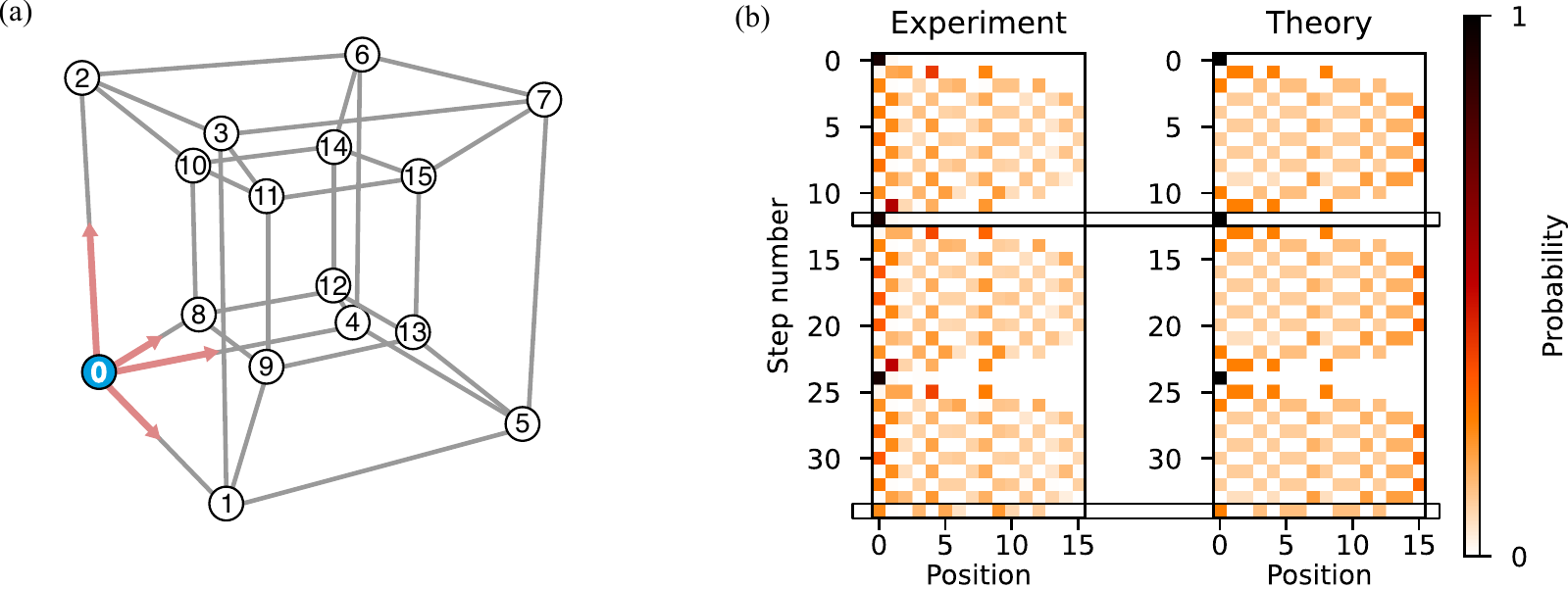}
	\caption{
        QW on a four-dimensional hypercube. 
        (a) We encode the nodes of the four-dimensional hypercubes as binary numerals, $0\ldots 2^4-1=15$.
        The walker is initialized at position $\Vec{x}=0$ in a coin-superposition state, with all four coin degrees of freedom uniformly populated.
        (b) Comparison between experimental data and theory predictions for the position distribution $P(\Vec{x},n)$ for different QW steps $n$.
        During each step, a four-dimensional Grover coin is applied to all positions. 
        Perfect state transfer occurs within a period of twelve steps. 
        % Horizontal boxes denote steps for which coin-resolved distributions are presented in (c) and (d).
        % (c) Coin-resolved probability for step 12.
        % Colored bars are measurement, grey empty bars are theory. 
        % After one twelve-step period, the walker returns to the initial configuration. 
        % Error bars comprise statistical errors and are too small to be easily visible. 
        % The four panels correspond to four coin states $c=0\ldots 3$.
        % (d) Same as (c) for step 34.
	}\label{fig:2}
\end{figure*}

\begin{figure*}[t]
	\includegraphics[width=\textwidth]{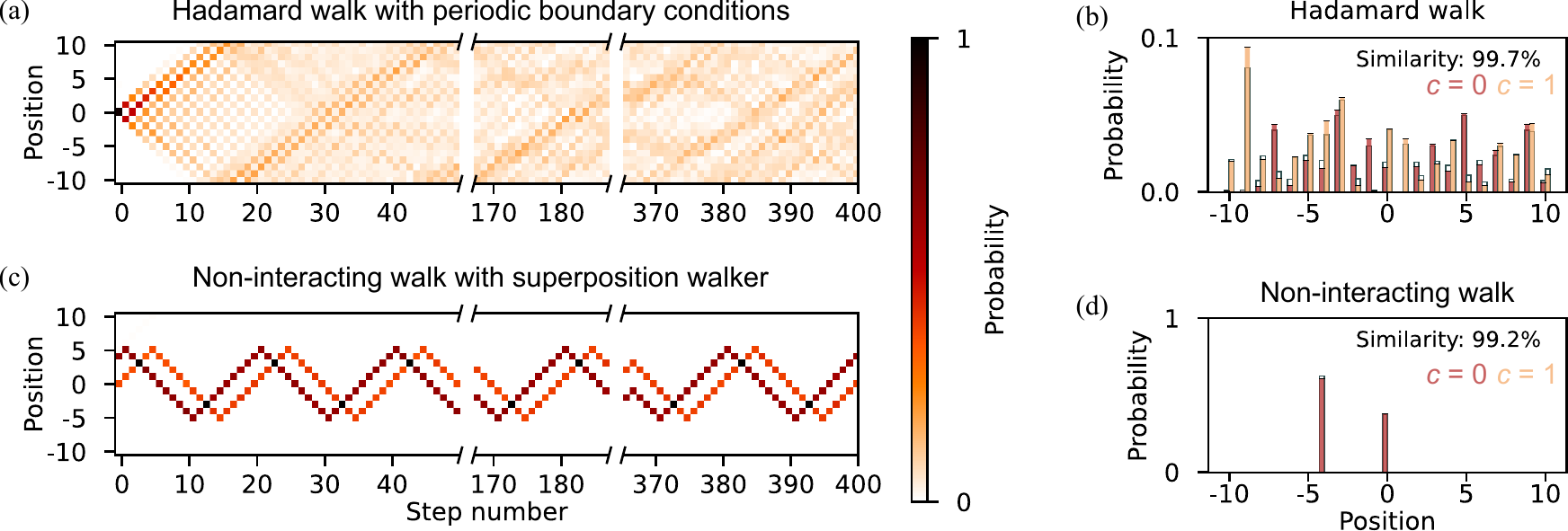}
	\caption{
        Bounded QWs over large step numbers.
        (a) Position distribution of a Hadamard QW on a circle with 21 positions. 
        We measure the evolution up to the 400th step (note skips on the horizontal axis) and see no degradation in similarities. 
        (b) Coin-resolved position distribution of the 400th step of the Hadamard walk. 
        We find a mean similarity of $\Bar{S}\approx 98\%$.
        Error bars indicate statistical uncertainties and are hardly visible.
        (c) Same as (a) but for a non-interacting walk with reflecting boundaries. 
        The walk was initialized with a walker in a position superposition. 
        (d) Same as (b) but for the non-interacting walk. 
        Because of the non-interacting nature of the walk, a population of the $c=1$ coin state is neither expected nor observed.
        The mean similarity yields $\Bar{S}\approx 99\%$. 
	}\label{fig:3}
\end{figure*}

\paragraph{Results.\textemdash\hspace*{-2ex}}
    We first implement Grover-type QW unitaries on four-dimensional hypercubes, an essential element for quantum computation based on discrete time QWs \cite{PhysRevA.81.042330} and notoriously hard to realize with other approaches due to the necessity for a four-dimensional coin with specific connectivity $(d_c=4)$ and their overall dimensionality of $d=64$. A Grover coin, $\hat{G}=2/d_c\sum_{c,c'=0}^{d_c-1}|c\rangle\langle c'|-\hat{1}$, encodes an equal superposition of all directions.
    Our initial state is localized at the origin with an equal coin superposition, $\ket{\psi} = 1/\sqrt{d_c}\sum_{c=0}^{d_c-1}|c\rangle\otimes|\Vec{x}=0\rangle$.
    In the experiment, we coherently distribute the amplitude of the OPO pulse over the four frequency bins associated with position $\Vec{x}=0$.
    We then calculate the unitary $\hat{U}$ corresponding to a QW with $N$ steps and measure the output probabilities $\{p\}$, from which we obtain the position distributions $P(\vec{x})$.
    Experimental results and the corresponding theory are shown in Fig. \ref{fig:2}(b).
    We compare measurement results and theory for each step via the Bhattacharyya coefficient, $S_n=\sum_{\Vec{x}}\sqrt{P_{\mathrm{ex.}}(\Vec{x}_n)P_{\mathrm{th.}}(\Vec{x}_n)}$, to quantify their similarities $S_n$.
    We also look at the mean similarity over $N$ steps, defined as $\Bar{S}=1/N\sum_{n=N}^T S_n$, ranging from $1$ for a perfect match for all times to $0$ for a complete mismatch.
    We find a high mean similarity, $\Bar{S}\approx 95\%$, which confirms that our system implements high-quality unitary operations.
    Note that the walker exhibits complete revival at the origin after the $12$th step, as expected from theory.
    More details as well as coin-resolved probability distributions can be found in the \cite{supp}.
    % This is also visualized in Fig. \ref{fig:2}(c), which illustrates that only $\Vec{x}=0$ is occupied in each coin-resolved distribution after step twelve.
    % In contrast, multiple positions become populated, e.g., after the $34$th step, Fig. \ref{fig:2}(d).
    % The mismatch between experiment and theory can be attributed to the imperfect mode-selectivity of our QPG.
    
    We next implement other interesting QW unitaries. 
    First, QWs with periodic boundary conditions, which are of significant interest (see, e.g., \cite{BEDNARSKA200321}) but are challenging to realize in an experiment because the leftmost and rightmost modes, which are typically spatially or temporally separated, must be coupled. 
    Only special cases have been demonstrated \cite{Bian17, Lorz19, Nejadsattari19} to date. 
    Second, QWs with superposition input states, which are central to many applications \cite{Shenvi03} but are challenging to realize in other platforms \cite{Su19} because coherent spatial or time-bin superposition states require intricate interferometers for their implementation.
    Again, more details are found in the \cite{supp}. 
    
    Figure \ref{fig:3} shows the results of two examples of QW on finite graphs\textemdash (a), (c): evolution of the position distributions $P(\Vec{x})$; (b), (d): coin-resolved distributions $p$ after the last step. 
    The results of Fig. \ref{fig:3}(a,b) correspond to implementing the unitary of a Hadamard walk over 400 steps with periodic boundaries at $\Vec{x}=\pm 10$\textemdash a circle with 21 nodes\textemdash and a localized input state $|c=0\rangle\otimes|\Vec{x}=0\rangle$.
    The dimensionality of the overall system is $d=42$.
    We find a high mean similarity of $\Bar{S}\approx 98\%$.
    We note that a comparable realization of this walk in the spatial domain would require roughly $4,500$ beam splitters and $22$ detectors.
    In Fig. \ref{fig:3} (c,d), we study an input superposition state \mbox{$|c=0\rangle\otimes\left(\sqrt{3/8}|\Vec{x}=-3\rangle+\sqrt{5/8}|\Vec{x}=0\rangle\right)$} that undergoes a non-mixing walk, exhibiting identity coin everywhere along with reflecting boundaries at $\Vec{x}=\pm 5$, again over several hundreds of steps. 
    Here, the dimensionality of the overall system is $d = 22$.
    In this case, we find a mean similarity of $\Bar{S}\approx 99\%$.

\paragraph{Discussion.\textemdash\hspace*{-2ex}}    
    We introduced a novel approach to implementing large, complex unitaries and benchmarked its performance by realizing unitaries associated with QWs.
    Our scheme does not suffer from error accumulation when realizing large unitaries.
    While conventional schemes require light to propagate through an ever increasing number of components, which incurs errors, our systems fidelity is only limited by the error in the QPG projection.
    This can be very small \cite{Ansari17}, which enables us to observe similarities of up to 99\%. 
    We note that, since our system does not suffer from geometric constraints, arbitrary modes can be interfered.
    This would enable the emulation of more exotic QWs, such as Levy flights \cite{BOUCHAUD1990127}, whereby each step moves the walker by several positions, or lazy walks, where the walker is allowed to stay at its position \cite{Childs:09}; these, however, are beyond the scope of the current work.

    The maximum number of FBs was limited to 64 by the bandwidth of the laser and the spectral resolution of the pump pulse shaper, as well as by the phase-matching bandwidth of the QPG: individual FBs must be at least as wide as the phase-matching to ensure high-fidelity projections \cite{Ansari18,Ansari17}.
    This number is on par with other approaches based on FBs \cite{PhysRevA.100.012320} and, with existing technology, can be scaled to more than 100 FBs. 
    % In addition, we demonstrated that there is hardly any limitation to the depth of the linear network, as witnessed by the high-quality realization of unitaries describing QWs with up to 400 steps.
    
    At this point, we should briefly discuss the impact of limited efficiency. 
    A simple QPG with an internal conversion efficiency of around 60\% has been demonstrated already \cite{Allgaier:17}, while efficiencies of close to 100\% have been shown with more complex QPGs \cite{Reddy:18}. 
    In this work, conversion efficiency was limited to a around 0.1\% (see \cite{supp}). 
    For the sake of this discussion, we assume a total efficiency (including coupling to and from the QPG and detection of the converted light) of 0.01\%. 
    We note that this number is always the same regardless of the implemented unitary and it must be compared to the accumulated losses when realizing the unitary with conventional approaches. 
    If we compare this to a state-of-the-art time-multiplexed QW with a step-to-step efficiency of 80\% \cite{PhysRevLett.125.213604}, our approach wins out after only 42 steps. 
    In fact, we would require a step-to-step efficiency of 97.7~\% to reach an overall efficiency of 0.01~\% after 400 steps. 
    While in principle doable, this number poses a challenging benchmark for experiments, which gets only more demanding the longer the evolution.

    In this work, the QPG was limited to a single output.
    This allows for measuring first-order output distributions, but is not sufficient for quantum applications that rely on higher-order correlations.
    However, this problem is readily solved by using a multi-output QPG \cite{PRXQuantum.4.020306}, which maps different input superpositions into different output frequencies.
    Such a device can ultimately realize large unitary operations while granting access to all output channels.

    Finally, we note that the existing QPG technology can be extended to achieve hundreds of modes, laying the foundation for realizing truly large-scale and highly interconnected quantum optical networks.
    Furthermore, we have already demonstrated that the QPG is compatible with genuine quantum light \cite{Allgaier:17, Ansari18prl, Ansari20}. 
    We believe that our approach can provide an advantage for frequency-encoded systems; it provides high-quality large-scale unitaries, compatibility with quantum light, and a favourable scaling behaviour, and can thus open a new pathway towards the implementation of photonic quantum computation.

\paragraph*{Acknowledgments.---\hspace*{-2ex}}

	We acknowledge financial support by the European Research Council through the ERC project QuPoPCoRN (Grant No. 725366).
	
	S. D. and V. A. contributed equally to this work. 

\bibliography{frequency_unitaries.bib}

\end{document}